\def\~{{$\tilde{\phantom{a}}$}}

\documentclass [12pt] {article}
\usepackage{epsfig}
\usepackage{color}

\def\thebibliography#1{\section{References}\markboth
 {REFERENCES}{REFERENCES}\list
 {[\arabic{enumi}]}{\settowidth\labelwidth{[#1]}\leftmargin\labelwidth
 \advance\leftmargin\labelsep
 \usecounter{enumi}}
 \def\newblock{\hskip .11em plus .33em minus -.07em}
 \sloppy
 \sfcode`\.=1000\relax}
\def\upcite#1{\raise6pt\hbox{\scriptsize
\cite{#1}}}
\def\lsim{\mathrel {\vcenter {\baselineskip 0pt \kern 0pt
    \hbox{$<$} \kern 0pt \hbox{$\sim$} }}}
\def\gsim{\mathrel {\vcenter {\baselineskip 0pt \kern 0pt
    \hbox{$>$} \kern 0pt \hbox{$\sim$} }}}
\def\gtlt{\mathrel {\vcenter {\baselineskip 0pt \kern 0pt
    \hbox{$>$} \kern 0pt \hbox{$<$} }}}

\def\hline{\noalign{\hrule \vskip2pt}}

\def\|{\ifmmode\Vert\else \char`\|\fi}
\ifx\oldzeta\undefined                          
  \let\oldzeta=\zeta                            
  \def\zzeta{{\raise 2pt\hbox{$\oldzeta$}}}     
  \let\zeta=\zzeta                              
\fi

\ifx\oldchi\undefined                           
  \let\oldchi=\chi                              
  \def\cchi{{\raise 2pt\hbox{$\oldchi$}}}       
  \let\chi=\cchi                                
\fi

\def\frac#1#2{{#1 \over #2}}

\def\half{\ifinner {\scriptstyle {1 \over 2}}
   \else {1 \over 2} \fi}

\def\ave#1{\left\langle#1\right\rangle} 

\def\simge{\mathrel{%
   \rlap{\raise 0.511ex \hbox{$>$}}{\lower 0.511ex \hbox{$\sim$}}}}
\def\simle{\mathrel{
   \rlap{\raise 0.511ex \hbox{$<$}}{\lower 0.511ex \hbox{$\sim$}}}}

\def\buildchar#1#2#3{{\null\!                   
   \mathop#1\limits^{#2}_{#3}                   
   \!\null}}                                    
\def\overcirc#1{\buildchar{#1}{\circ}{}}

\def\slashchar#1{\setbox0=\hbox{$#1$}           
   \dimen0=\wd0                                 
   \setbox1=\hbox{/} \dimen1=\wd1               
   \ifdim\dimen0>\dimen1                        
      \rlap{\hbox to \dimen0{\hfil/\hfil}}      
      #1                                        
   \else                                        
      \rlap{\hbox to \dimen1{\hfil$#1$\hfil}}   
      /                                         
   \fi}                                         %

\def\subrightarrow#1{
  \setbox0=\hbox{
    $\displaystyle\mathop{}
    \limits_{#1}$}
  \dimen0=\wd0
  \advance \dimen0 by .5em
  \mathrel{
    \mathop{\hbox to \dimen0{\rightarrowfill}}
       \limits_{#1}}}                           

\def\overlay#1#2{\ifmmode%
\setbox0=\hbox{$#1$}%
\setbox1=\hbox to\wd0{\hss$#2$\hss}\else%
\setbox0=\hbox{#1}%
\setbox1=\hbox to\wd0{\hss#2\hss}\fi%
#1\hskip-\wd0\box1 }

\def\pmb#1{\leavevmode\setbox0=\hbox{#1}%
\kern-.02em\copy0\kern-\wd0
\kern.04em\copy0\kern-\wd0
\kern-.02em\raise.04em\box0 }

\def\vereq#1#2{\lower3pt\vbox{\baselineskip1.5pt \lineskip1.5pt
\ialign{$\m@th#1\hfill##\hfil$\crcr#2\crcr\sim\crcr}}}

\def\tensor#1{\protect\@ontopof{#1}{\leftrightarrow}{1.15}\mathord{\box2}}
\def\overstar#1{\protect\@ontopof{#1}{\ast}{1.15}\mathord{\box2}}
\def\overdots#1{\protect\@ontopof{#1}{\cdots}{1.0}\mathord{\box2}}
\def\overcirc#1{\protect\@ontopof{#1}{\circ}{1.2}\mathord{\box2}}
\def\loarrow#1{\protect\@ontopof{#1}{\leftarrow}{1.15}\mathord{\box2}}
\def\roarrow#1{\protect\@ontopof{#1}{\rightarrow}{1.15}\mathord{\box2}}

\def\@ontopof#1#2#3{%
{\mathchoice
{\@@ontopof{#1}{#2}{#3}\displaystyle\scriptstyle}%
{\@@ontopof{#1}{#2}{#3}\textstyle\scriptstyle}%
{\@@ontopof{#1}{#2}{#3}\scriptstyle\scriptscriptstyle}%
{\@@ontopof{#1}{#2}{#3}\scriptscriptstyle\scriptscriptstyle}%
}%
}

\def\@@ontopof#1#2#3#4#5{%
\setbox0=\hbox{$#4#1$}%
\setbox1=\hbox{$#5#2$}%
\setbox2=\hbox{}\ht2=\ht0 \dp2=\dp0 %
\ifdim\wd0>\wd1 %
\setbox1=\hbox to\wd0{\hss\box1\hss}%
\mathord{\rlap{\raise#3\ht0\box1}\box0}%
\else   %
\setbox1=\hbox to.9\wd1{\hss\box1\hss}%
\setbox0=\hbox to\wd1{\hss$#4\relax#1$\hss}%
\mathord{\rlap{\copy0}\raise#3\ht0\box1}%
\fi
}%

\def\lambdabar{\protect\@lambdabar}
\def\@lambdabar{%
\relax
\bgroup
\def\@tempa{\hbox{\raise.73\ht0
\hbox to0pt{\kern.25\wd0\vrule width.5\wd0
height.1pt depth.1pt\hss}\box0}}%
\mathchoice{\setbox0\hbox{$\displaystyle\lambda$}\@tempa}%
{\setbox0\hbox{$\textstyle\lambda$}\@tempa}%
{\setbox0\hbox{$\scriptstyle\lambda$}\@tempa}%
{\setbox0\hbox{$\scriptscriptstyle\lambda$}\@tempa}%
\egroup
}

\def\corresponds{{\lower.2ex\hbox{=}}{\rm\kern-.75em^\triangle}}
\def\succsim{\succ\kern-.9em_\sim\kern.3em}
\def\precsim{\prec\kern-1em_\sim\kern.3em}
\def\slantfrac#1#2{\kern1em^{#1}\kern-.3em/\kern-.1em_{#2}}

\begin{document}
                                                                
\begin{center}
{\Large\bf Electron Bubbles in Liquid Helium}
\\

\medskip

Kirk T.~McDonald
\\
{\sl Joseph Henry Laboratories, Princeton University, Princeton, NJ 08544}
\\
(November 12, 2000)
\end{center}

\section{Problem}

When an electron (or positronium atom) is injected into liquid helium with
nearly zero energy, a bubble quickly forms around it.  This phenomenon (which
also occurs in liquid hydrogen, liquid neon and possibly in solid helium)
lowers the mobility of the
electron to a value similar to that for a positive ion.

Estimate the radius of the bubble at zero pressure and temperature.

If the liquid is held in a state of negative pressure, the bubble will
expand beyond the radius at zero pressure.  Estimate the negative pressure
such that a bubble once formed will grow without limit.

\section{Solution}

\subsection{Bubble Radius at Zero Pressure}

For a simple estimate, we follow the original paper by R.A.~Ferrell
\cite{Ferrell}.  We assume the bubble to be spherical with radius $a$.

Our estimate is based on an energy argument.  The bubble is kept from 
collapsing by the pressure due to the collisions of the electron with the wall.
At zero temperature, the motion of the electron inside the bubble is due to
zero-point energy.  We relate this to the zero-point momentum, which we
estimate via the uncertainty principle:
\begin{equation}
\delta x\ \delta p_x \approx {\hbar \over 2}.
\label{s1}
\end{equation}
For an electron inside a bubble of radius $a$, the uncertainty in coordinate
$x$ is about $2 a / 3$, so we estimate that
\begin{equation}
\ave{p_x^2}^{1/2} \approx \delta p_x \approx {3 \hbar \over 4 a}.
\end{equation}
The zero-point energy is therefore estimated as
\begin{equation}
U_{\rm zero-point} \approx {\ave{p^2} \over 2 m}
\approx {3\ave{p_x^2} \over 2 m}
\approx {27 \hbar^2 \over 32 a^2 m}
\approx {\hbar^2 \over a^2 m}.
\label{s2}
\end{equation}
[The formal result for the zero point energy for a particle of mass $m$ inside
an infinite spherical potential well of radius $a$ is 
$\pi^2 \hbar^2 / 2 a^2 m$.]

The bubble tends to collapse to zero due to the force of surface tension.
We characterize surface tension by a coefficient $\gamma$ which is a force
per unit length = energy per unit area.  A bubble of radius $a$ has a 
surface energy given by
\begin{equation}
U_{\rm surface} = 4 \pi a^2 \gamma.
\label{s3}
\end{equation}

The physical radius of the bubble minimizes the total energy, so we solve
\begin{equation}
0 = {dU_{\rm total} \over da} = - {2 \hbar^2 \over a^3 m} + 8 \pi a \gamma,
\label{s4}
\end{equation}
which yields
\begin{equation} 
a_{\rm bubble} \approx \left( {\hbar^2 \over 4 \pi \gamma m} \right)^{1/4}
= \left( {(\hbar c)^2 \over 4 \pi \gamma m c^2} \right)^{1/4}.
\label{s5}
\end{equation}

To estimate the surface tension coefficient $\gamma$ for liquid helium,
we recall that it liquifies at 4K where the kinetic energy per atom is
about 1/4000 eV.  So, we suppose that the
binding of a helium atom to its neighbors on the surface is about 1/4000 eV, and
that the distance between atoms is about one Angstrom.  Hence,
\begin{equation}
\gamma \approx {2.5 \times 10^{-4}\ {\rm eV} \over (10^{-8}\ {\rm cm})^2}
= 2.5 \times 10^{12}\ {\rm eV/cm}^2
= 4\ {\rm erg/cm}^2.
\label{s6}
\end{equation}
[Apparently, this estimate is about a factor of ten high.]

We also recall that the electron rest energy is $m c^2 \approx 5 \times 10^5$
eV, and that $\hbar c \approx 200$ MeV-fermi $= 2 \times 10^{-5}$ eV-cm.
Hence, we estimate that
\begin{equation} 
a_{\rm bubble} \approx 
\left( {(2 \times 10^{-5})^2 \over 4 \pi \cdot 2.5 \times 10^{12} \cdot 
5 \times 10^5}  \right)^{1/4}\ {\rm cm}
\approx \left( {10^{-28} \over 4} \right)^{1/4}\ {\rm cm}
\approx 7\ {\rm \AA}. 
\label{s7}
\end{equation}
The experimental value for the bubble radius $a$ is about 17 \AA.  [Noting
that our estimate for $U_{\rm zero-point}$ is a factor of five low and that
for $\gamma$ is a factor of ten high, the estimate (\ref{s7}) should be 
multiplied by $\sqrt[4]{50}$ to yield a prediction that $a_{\rm bubble} \approx
19$ \AA.]

\subsection{Negative Pressure}

When a bubble is formed in a liquid at positive pressure $P$, additional
work $PV$ must be done.  We consider this work to be stored in the bubble in
the form of an energy, and so the total energy of a bubble of radius $a$ is
\begin{equation}
U = U_{\rm zero-point} + U_{\rm surface} +  PV 
\approx {C \hbar^2 \over a^2 m} + 4 \pi a^2 \gamma + {4 \pi a^3 P \over 3},
\label{s8}
\end{equation}
where $C = 1$ in our approximate model, and $C = \pi^2 / 2$ from a
calculation based on a deep spherical potential well.

If the bubble is to grow indefinitely we must have $\partial U / \partial a
< 0$ for all $a$.
The zero-point energy term decreases monotonically with radius $a$, while the
surface energy term increases with radius.  Together, the first two terms
have a single minimum at the radius found in sec.~2.1.

\begin{figure}[htp]  
\begin{center}
\includegraphics[width=4in, angle=0, clip]{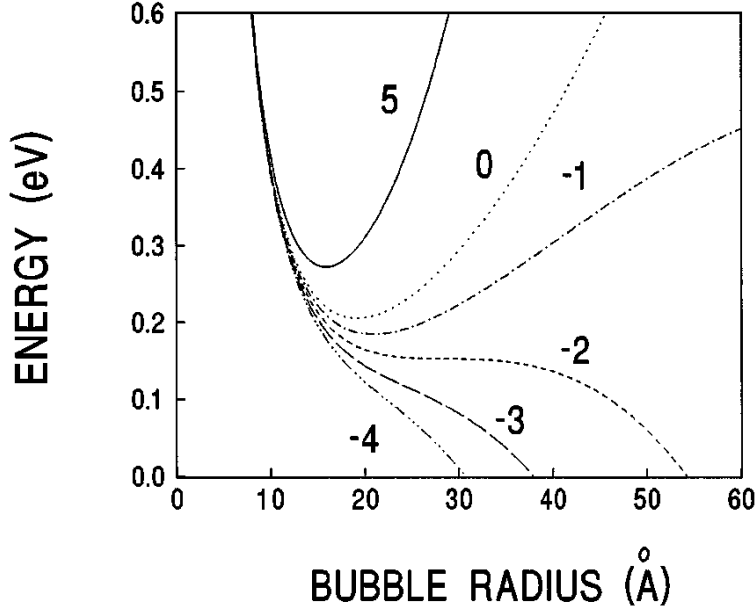}
\parbox{5.5in} 
{\caption[ Short caption for table of contents ]
{\label{Marisfig} The energy (\ref{s8}) of an electron bubble in liquid
helium at zero temperature as a function of the radius $s$.  The curves are
labeled by the pressure in bars \cite{Maris96}. 
}}
\end{center}
\end{figure}

For negative pressures, the pressure term decreases with radius more rapidly
than the surface term increases with radius.  So at large radii the
negative pressure term dominates, and $\partial U / \partial a$ becomes
negative.  For small negative pressures there is a maximum of $U$ at a
radius larger than that of the minimum of $U$, and between these two
extrema the slope is positive.  As the magnitude of the negative pressure
increases, the two extrema approach one another, until at the desired
critical pressure they coalesce, and there is a radius for which both
$\partial U / \partial a = 0$ and $\partial^2 U / \partial a^2 = 0$, as
illustrated in Fig.~\ref{Marisfig}

From eq.~(\ref{s8}), the critical pressure and radius are related by
\begin{eqnarray}
{\partial U \over \partial a} & = & 0 = - {2 C \hbar^2 \over a^3 m} 
+ 8 \pi a \gamma + 4 \pi a^2 P,
\label{s9} \\
{\partial^2 U \over \partial a^2} & = & 0 =  {6 C \hbar^2 \over a^4 m} 
+ 8 \pi \gamma + 8 \pi a P.
\label{s10}
\end{eqnarray}
We quickly find that
\begin{equation}
a = \left( {5 C \hbar^2 \over 4 \pi \gamma m} \right)^{1/4}
= 5^{1/4} a_{\rm bubble}(P=0)
\approx 10 C^{1/4} \ {\rm \AA}.
\label{s11}
\end{equation}
and
\begin{equation}
P = - {8 \gamma \over 5 a}
\label{s12}
\end{equation}
Using our estimate (\ref{s6}) that $\gamma \approx 0.004$ N/m and taking
$C = 1$, we find the critical pressure to be $P = - 64$ bar.
This is about 40 times the reported value of -1.6 bars \cite{Maris96}.  
Since the result
(\ref{s12}) varies as $\gamma^{5/4}$, it is more sensitive to the value
of surface tension than is the bubble radius.  If we use $\gamma = 0.0004$ N/m
and $C = \pi^2/2$, then our estimate for the critical negative pressure would
be -2.4 bars.

\end{document}